# Experimental setup for low energy laser-based angle resolved photoemission spectroscopy

J. D. Koralek[1,2], J. F. Douglas[1], N. C. Plumb[1], J. D. Griffith[1], S. T. Cundiff[2],

H. C. Kapteyn[1,2], M. M. Murnane[1,2], and D. S. Dessau[1,2]

[1]Department of Physics, University of Colorado, Boulder, CO 80309-0390, USA,

[2]JILA, University of Colorado and NIST, Boulder, CO 80309-0440, USA,

A laser-based angle resolved photoemission (ARPES) system utilizing 6 eV photons from the fourth harmonic of a mode-locked Ti:Sapphire oscillator is described. This light source greatly increases momentum resolution and photoelectron count rate, while reducing extrinsic background and surface sensitivity relative to higher energy light sources. In this review, the optical system is described, and special experimental considerations for low energy ARPES are discussed. The calibration of the hemispherical electron analyzer for good low-energy angle-mode performance is also described. Finally, data from the heavily studied $Bi_2Sr_2CaCu_2O_{8+\delta}$ (Bi2212) high $T_c$ superconductor is compared to results from higher photon energies.

# 1. Introduction

Angle resolved photoemission spectroscopy (ARPES) has become a key tool in the study of the electronic structure of solids.[1] The technique is based on Einstein's photoelectric effect, where photons of sufficient energy eject electrons from a solid. Since momentum is conserved in this process, the angular distribution of photoelectrons from a single crystal is representative of the momentum distribution of initial electronic states in that crystal. With the ability of modern electron spectrometers to analyze electrons from many angles at once, ARPES is like taking a snapshot of the electrons in momentum space. Significant improvements in ARPES technology over the past decade have allowed researches to study not only band structure in simple materials, but also to learn about the electron-electron interactions in strongly correlated systems such as the high $T_c$ cuprate superconductors.[2,3]

Standard ARPES experiments are carried out using discharge lamps or synchrotron light sources operating in the 20 - 100 eV photon energy range. This photon energy range is at the universal minimum of photoelectron mean free path in solids[4] (Figure 1), meaning that the resulting ARPES spectra may be heavily influenced by surface physics. Longer mean free paths are therefore desirable for the study of bulk physics such as superconductivity. Attempts have been made to increase bulk sensitivity using very high photon energy,[5,6] but these systems typically suffer from low count rates and poor resolution.

Past work on using lasers for photoemission includes 2-photon photoemission,[7,8,9] combined laser and synchrotron pump-multiple-probe photoemission,[10] high harmonic generation based light sources[11,12] and ultrahigh energy-resolution photoemission.[13] However, none of these laser-based systems have been able to perform angle-resolved measurements in which band dispersion and self-energy effects can be studied. In this review we describe a 6 eV laser-based photoemission system capable of true angle-resolved measurements, which has recently been used to produce the clearest images yet of electrons in a high $T_c$ superconductor.[14] Working at this photon energy increases bulk-sensitivity by about an order of magnitude compared to standard ARPES (Figure 1), though the actual surface sensitivity of course varies somewhat from material to material depending on the plasma frequencies, etc.. The low photon energy also greatly increases the momentum resolution since the electronic states are more widely dispersed in angle. Compared to synchrotron sources, there is also a sizeable reduction in operating costs and space requirements which should open up the ARPES technique to researchers without access to synchrotron beam time. Also exciting is the possibility to perform dynamic pump-probe ARPES measurements using the pulsed nature of the Ti:Sapphire laser.

In section II of this review we describe the 6 eV laser system in detail. Section III is dedicated to discussion of the UHV ARPES chamber. Section IV focuses on the electron analyzer and calibration for use at very low kinetic energy. In section V we show laser ARPES data from the high $T_c$ superconductor Bi2212 and compare with results from higher photon energies.

## II.  The 6 eV laser system

The optical layout for the 6 eV laser system is sketched in figure 2. At the heart of the system is a Kerr mode-locked Ti:Sapphire oscillator pumped with 5 Watts from a frequency doubled Nd:Vanadate laser. The oscillator generates 70 fs, 6 nJ pulses tunable around 840 nm (~1.5 eV) at a repetition rate of 100 MHz. Although much higher pulse energies are available from amplified systems, we use high repetition-rate and low pulse energy to avoid possible space charge complications.[15] In order to produce ultraviolet photons, we utilize 2 stages of non-linear second harmonic generation through type I phase matching in β-Barium Borate (BBO)[16]. The pulses are first focused into a 2 mm thick BBO crystal cut at the angle of 29.2° generating about 150 mW of the 2nd harmonic at 420 nm (~3 eV). The 2nd harmonic is separated from the fundamental with dichroic mirrors and then focused into a 1mm thick BBO crystal cut at 67.2° resulting in generation of the 4th harmonic at 210 nm (~6 eV). The 2 beams are then separated using either prisms or dichroic mirrors (shown in fig. 2).

The power of the 4th harmonic varies up to about 1 mWatt depending on the wavelength being used. Fourth harmonic power decreases with increasing photon energy as the absorption edge of the BBO crystal is approached, a property which can vary from crystal to crystal. Currently, the maximum usable photon energy achieved with reasonable flux is around 6.05 eV. Figure 3(a) shows a typical 4th harmonic spectrum along with a Gaussian fit with a full-width half-max of 4.7 meV. The power for this spectrum was measured to be 200 µW which corresponds to $2 \times 10^{14}$ photons/s (we have achieved flux of up to $10^{15}$ photons/s). This combination of high flux and narrow bandwidth represents a considerable improvement over even the best undulator beamlines at synchrotron facilities.

After separation from the 2$^{nd}$ harmonic, the 4$^{th}$ harmonic passes through a rotatable ½-wave plate, allowing any linear polarization to be used for ARPES. This ease of polarization control is an important advantage of the laser system since the ARPES signal/background depends heavily on polarization through the ARPES matrix elements.[1,2,3] A ¼ wave plate can also be used to obtain circularly polarized light, which can be used for studies of magnetism or to test for time-reversal breaking effects[17,18] Finally, the ultraviolet light is focused into the UHV chamber though a MgF viewport, using a curved Al mirror. The curved mirror is mounted on a linear translation stage so that the spot size on the sample can be adjusted. This combination of high flux, improved resolution, low operating cost and ease of polarization and focus control, make lasers an excellent light source for ARPES.

## III. UHV system

The laser ARPES chamber is constructed of 316 stainless steel, and typically maintains a base pressure below $2 \times 10^{-11}$ Torr after bakeout. Samples are introduced into the vacuum through a 2-stage load-lock transfer system and are measured on a 5-axis He cooled cryostat capable of temperatures as low as 8.5 K. Samples share a common ground with the electron analyzer. The 6 eV photons enter the chamber at an angle of 45 degrees relative to the entrance of the hemispherical electron analyzer as sketched in figure 2. The entrance slit of the electron analyzer is perpendicular to the schematic in figure 2. Among the other tools on the UHV chamber are an x-ray source and monochromator for XPS surface analysis, and a He discharge lamp for 21.2 eV ARPES.

Between the time photoelectrons leave the sample and are measured at the detector, they are subject to interactions that can alter their trajectory as they travel through the UHV chamber. The very low kinetic energy photoelectrons in laser ARPES have longer mean free paths in the sample, but are more susceptible to deflection by stray fields than those in high energy ARPES. Insulating objects in the ARPES chamber that are hit with electrons will charge, causing electric fields that can deflect the photoelectrons on their way to the analyzer. Therefore, any insulating material in the chamber is shielded with metal that is connected to ground. To minimize magnetic fields, only non-magnetic material is used on the sample manipulator and all heaters are counter-wound. Pumps and motors outside the chamber are kept away from the analyzer and sample positions. The entire chamber is shielded by two isolated layers of 0.04 inch thick annealed μ-metal, which are contiguous with the μ-metal shields of the hemispherical electron analyzer. With this shielding the magnetic field measured at the sample position is about 1 mGauss in magnitude.

## IV. The electron analyzer

One advantage of low energy ARPES is the improved **k**-resolution due to the favorable angle to **k** conversion. Unfortunately, obtaining good angular resolution at low energy is more challenging because of the increased sensitivity to stray fields. While the standard calibration procedures used at high energy are sufficient to achieve good energy resolution with 6 eV photons, additional steps must be taken to achieve good angular resolution.

The laser ARPES system uses a Scienta[19] SES 2002 hemispherical analyzer, whose main physical modification is the mating of the µ-metal shields of the lens to those of the UHV chamber. This was necessary to achieve fields of 1 mGauss in the sample analysis position. The combined energy resolution of the system can be seen in figure 3, which shows a raw photoemission spectrum from polycrystalline gold taken with 6 eV photons at T = 20 K. The data was taken at 1 eV pass energy using a 0.2 mm curved analyzer entrance slit. The Fermi-function fit has a 10% - 90% width of 10.7 meV, which is dominated by the sample temperature. We observe no difference in the Fermi widths from freshly evaporated gold versus gold transferred directly from atmospheric pressure, an indication that laser ARPES is indeed bulk-sensitive.

In order to achieve good angle-mode performance at low kinetic energy, the Scienta electron-lens voltage tables were constructed empirically for each pass energy. This was done through an iterative process, using angular calibration devices that were well characterized at higher kinetic energy. Figure 4 shows one such device, consisting of a small gold post and an array of thin wires. Electrons are photoemitted from the face of the gold post, and a shadow pattern is generated by the wire array as it blocks some of the electrons on their way to the analyzer. The gold sample and the wire array are both mounted on a rotatable stage which transfers into the cryostat in exactly the same manner as real samples. The dimensions of the device were verified using an optical microscope and a translation stage equipped with position encoders. The device was further tested by ARPES using 21.2 eV photons from a He discharge lamp, indeed confirming the predicted shadow pattern. It is critical that the device be evenly coated with colloidal graphite (except on the gold post) to minimize the effects of electron bombardment and work function variation across the surface.

The calibration procedure involved smoothly modifying the voltage tables, primarily for cylindrically symmetric lenses L6, L7, and L8[20], in order to make the ARPES data match the predicted pattern without dispersing in energy. Figure 5 shows calibration data collected using 6 eV photons, with an analyzer pass energy of 1 eV in swept mode. We estimate the angular resolution of the instrument to be $\pm 0.16°$ based on a convolution of the expected calibration pattern with a Gaussian. This angular resolution is maintained all the way down to 600 meV kinetic energy. Although better angular resolution has occasionally been achieved at higher kinetic energy, the favorable angle to **k** conversion at 6 eV yields a **k** resolution of $\pm 0.0018$ Å$^{-1}$ for a sample work function of 4.5 eV. To our knowledge this is the highest momentum resolution ever obtained from an electron spectrometer, and has allowed us to measure exceptionally sharp spectra from real materials, as presented in the next section.

## V. Comparison of results from Bi2212

We have recently reported 6 eV laser ARPES showing the sharpest images yet of the electronic structure of the high $T_c$ superconductor Bi2212.[14] This is perhaps the material most heavily studied by high resolution ARPES,[2,3] so it is important to compare 6 eV results with those at higher photon energies. Figure 6 shows ARPES data from optimally doped Bi2212 taken with the (a) 6 eV laser photons, (b) 28 eV photons from beamline 12.0.1 at the Advanced Light Source (ALS), and (c) 52 eV photons from ALS beamline 10.0.1. The energy distribution curves (EDCs) and momentum distribution curves (MDCs) at the Fermi surface from these data are compared in figure 7. A constant offset was subtracted from (c) to account for the 2$^{nd}$ order light from the monochromator (note that this does not take into account any structure in the photoemission from 2$^{nd}$ order light). A Mg filter was used to suppress 2$^{nd}$ order light for the data of panel (b). The band dispersion, derived from fits to MDCs, for the 6 eV data (open red circles) is shown on all 3 plots for direct comparison. The dispersion for panels (b) and (c), blue squares and black triangles respectively, are in very good agreement with the 6 eV data. Especially important is the fact that the dispersion kink[2] at roughly 70 meV is clearly reproduced at low photon energy. As discussed in ref 14, this is an indication that the low energy photoelectrons are still in the sudden limit[1,2,21,22,23] with respect to the interactions causing this abrupt change in the dispersion. In other words, the superconductor has not relaxed via this excitation channel prior to the photoelectron leaving the sample. This allows one to study the self energy associated with this interaction, which may play an important role in high $T_c$ superconductivity.

Although ARPES at 6 eV appears to be in the sudden limit relative to the nodal kink, it could be in the adiabatic limit compared to some higher energy excitations. However, since the peaks sharpen symmetrically as photon energy is lowered (the dispersions of figure 6 are identical), we deduce that at this temperature and resolution, we cannot see any peak sharpening resulting from a breakdown of the sudden approximation.[14] Such a sharpening (where spectral weight is transferred from within a peak) would be asymmetric since spectral weight is always transferred to lower binding energy as the adiabatic regime is approached. This transfer of incoherent spectral weight could contribute to the reduction in high energy background observed in figure 7(a), however, the reduced inelastic portion of the background expected at low energy will also significantly contribute to the observed difference in the EDCs. In the 20 - 50 eV photon energy range, electrons from many Brillouin zones are excited, all of which can scatter inelastically and contribute to the extrinsic background. At 6 eV, only electrons from about 2/3 of the 1$^{st}$ zone are excited, greatly reducing this extrinsic portion of the photoemission spectrum relative to the signal of interest.

Perhaps the most noticeable difference between the ARPES data of figure 6 (and figure 7) is the increased sharpness of the 6 eV data, for which there are several contributing factors. The most significant contribution is the improved momentum resolution at low photon energy, which not only decreases the MDC width (Figure 7a), but also decreases the EDC width (figure 7b) for these highly dispersive nodal states.[14] With this reasoning one would expect the 28 eV data to be sharper than the data from 52 eV, yet the opposite is observed. This is due to the bi-layer splitting, or doubling of the nodal band in Bi2212 due coupling between the $CuO_2$ planes within a unit cell.[24] At 52 eV, the ARPES matrix elements heavily suppress the anti-bonding band, while at 28 eV both bands are present with similar intensity, but are not independently resolved. At 6 eV the bonding band is essentially completely suppressed[25]. Another possible contribution to the sharp peaks observed at 6 eV is the increased bulk-sensitivity. Although a change in the underlying electronic structure is not observed, extrinsic surface effects, such as scattering from surface contaminants, will be reduced in the more bulk-sensitive experiments. Related to the increased bulk sensitivity are increased final-state lifetimes for low photon energy ARPES. This implies a reduced integration over $k_\perp$, meaning less extrinsic broadening for all but perfectly two-dimensional samples[26]. This has been shown to be an important effect even for naturally layered systems such as the cuprate superconductors.[27]

## VI. Summary

We have developed a state of the art ARPES system using 6 eV photons from a Ti:Sapphire laser. Through extensive shielding of the UHV chamber, and careful calibration of the electron analyzer, we have been able to overcome the difficulties associated with low energy photoelectron measurement, resulting in a significant increase in momentum resolution. The laser also offers increased photon flux at reduced bandwidth relative to higher energy light sources. The decreased surface sensitivity of low energy ARPES make it an indispensable tool for the study of bulk physics, possibly extending the technique to materials that do not cleave. These advantages point towards a promising future for lasers as a photoemission light source.

# Acknowledgements

Primary support for this work came from DOE grant DE-FG02-03ER46066, with other funding from NSF grants DMR 0402814 and DMR 0421496, and by the NSF ERC for Extreme Ultraviolet Science and Technology under NSF Award No. 0310717. We thank Y. Aiura, H. Eisaki, and K. Oka for growing the Bi2212 samples. We also thank M. Bunce, E. Erdos, A. V. Fedorov, T.J. Reber, M. Thorpe, and Q. Wang for assistance.

[1] S. Hüfner, *Photoelectron Spectroscopy* (Springer, Berlin 1995).

[2] A. Damascelli, Z. Hussain, and Z. -X. Shen, Rev. Mod. Phys. **75**, 473 (2003).

[3] Z.-X. Shen and D. S. Dessau, Phys. Rep. **253**, 1 (1995).

[4] M. P. Seah and W. A. Dench, Surf. Interface. Anal. **1**, 2 (1979).

[5] P. Torelli *et al.*, Rev. Sci. Inst., **76**, 023909 (2005)

[6] Also see many papers in "Proceedings of the Workshop on Hard X-ray Photoelectron Spectroscopy – HAXPES" 547, Issue 1, pages 1-238 (July 2005)

[7] R. Haight *et al.* Rev. Sci. Inst. **59**, 1941(1988)

[8] W. Nessler *et al*, Phys. Rev. Lett. **81**, 4480-4483 (1998)

[9] Y. Sonoda and T. Munakata, Phys. Rev. B **70**, 134517 (2004)

[10] T. Gießel *et al.,* Rev. Sci. Inst. **74**, 11 (2003)

[11] P. Siffalovic *et al.*, Rev. Sci. Inst. **72**, 1 (2001)

[12] M. Bauer *et al.,* Phys. Rev. Lett. **87**, 025501 (2001)

[13] T. Kiss *et al.*, Phys. Rev. Lett. **94**, 057001 (2005).

[14] J. D. Koralek *et al.,* Phys. Rev. Lett., **96**, 017005 (2006)

[15] X. Zhou *et al.*, J. Electron Spectrosc. Relat. Phenom. **142**, 27 (2005)

[16] R. L. Sutherland, *Handbook of Nonlinear Optics* (Marcel Dekker, New York 2003)

[17] A. Kaminski *et al.*, Nature, **416**, 610 (2002)

[18] S. Borisenko *et al,*, Phys. Rev. Lett. **96**, 067001 (2006)

[19] Brand name given for technical reasons only and does not constitute an endorsement by *NIST*.

[20] Gammadata Scienta AB, *SES 2002 Manual* (2001)

[21] L. Hedin and J. D. Lee, J. Electron Spectrosc. Relat. Phenom. **124**, 289 (2002).

[22] J. Stohr *et al.*, Phys. Rev. Lett. **51**, 821 (1983).

[23] J. D. Lee, O. Gunnarsson, and L. Hedin, Phys. Rev. B **60**, 8034 (1999).

[24] A. A. Kordyuk *et al*. Phys. Rev. B **70**, 214525 (2004)

[25] A. Ino *et al*., unpublished

[26] N. V. Smith *et al*. Phys. Rev. B **47**, 15476 (1993).

[27] S. Sahrakorpi *et al*. Phys. Rev. Lett. **95**, 157601 (2005)

FIG. 1. (Color online) The "universal curve" for surface sensitivity in photoemission.[4] Electron inelastic mean free paths from a variety of materials are plotted versus kinetic energy relative to $E_F$ (the lowest kinetic energies shown will not be able to overcome the work function). Indicated on the plot are the kinetic energy ranges for standard ARPES and laser ARPES.

FIG. 2. (Color) Schematic layout of the 6 eV laser system. All lenses have 5 cm focal length and are UV fused silica where necessary. Dichroic mirrors are shown in dark green.

FIG. 3. (Color online) (a) Typical 4$^{th}$ harmonic spectrum (open circles) and Gaussian fit (Solid line). (b) Raw photoemission spectrum from polycrystalline gold at T = 20 K (open circles) and Fermi-Dirac function fit (solid line).

FIG. 4 (Color) Rotatable ARPES calibration device. Electrons are photoemitted from the face of the small gold post and a shadow pattern is generated as some electrons are blocked by the wire array on their way to the analyzer. The blue line represents light incident on the gold sample.

FIG. 5 (Color) ARPES image of the angular calibration device taken using 6 eV laser photons.

FIG. 6. (Color) Comparison of nodal ARPES data from optimally doped Bi2212 taken with (a) 6 eV photons at T = 25 K, (b) 28 eV photons at T = 26 K, and (c) 52 eV photons at T = 16 K. The MDC derived dispersion for the 6 eV data is shown on all three panels (open red circles), and the dispersion for the 28 eV (blue squares) and 52 eV (black triangles) data are on panels (b) and (c) respectively. Because of the high flux of the laser, the data of panel (a) was acquired in less than 2 minutes.

FIG. 7. (Color) Comparison of EDCs at $\mathbf{k}_F$ (a) and MDCs at $E_F$ for the data of figure 6.

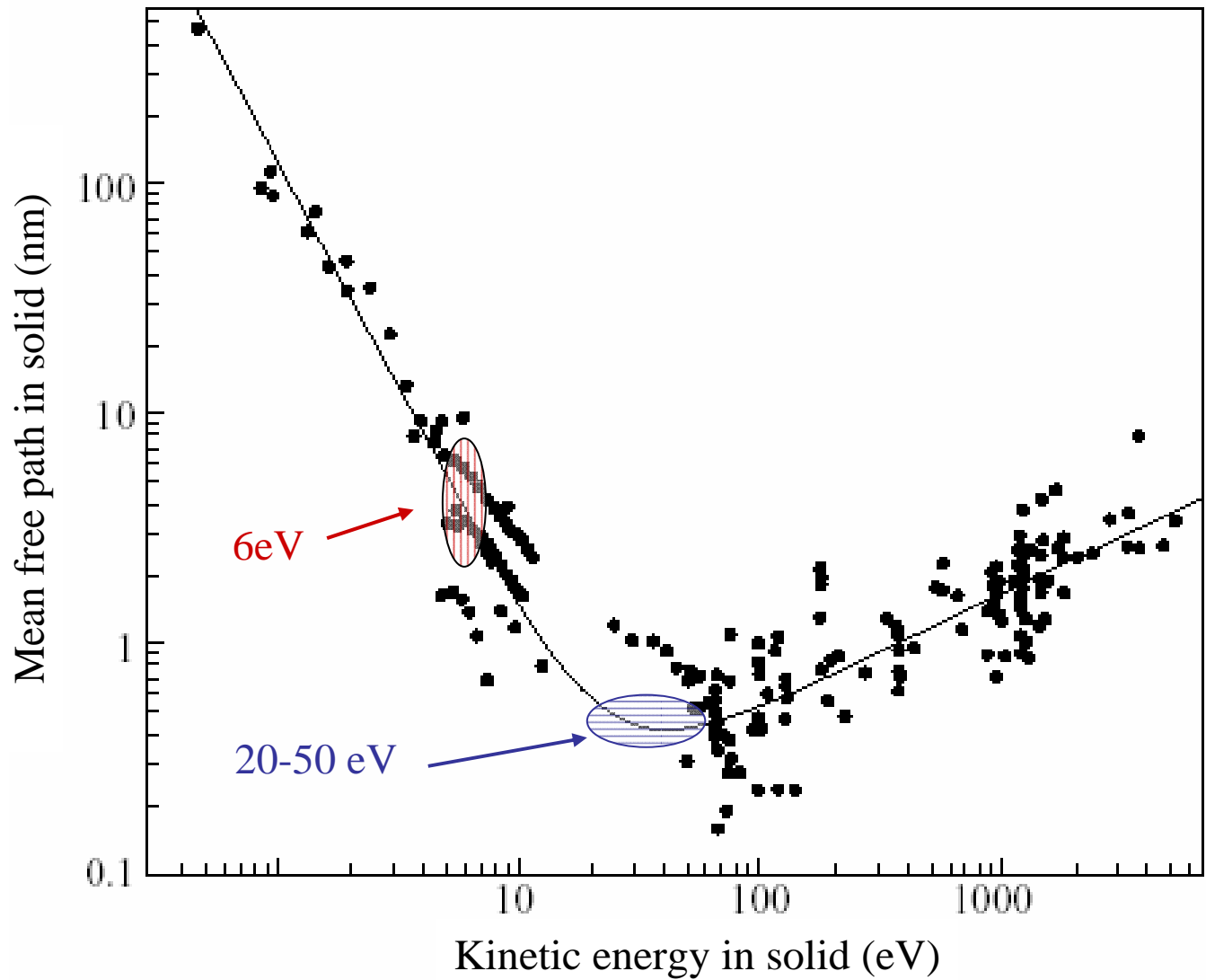

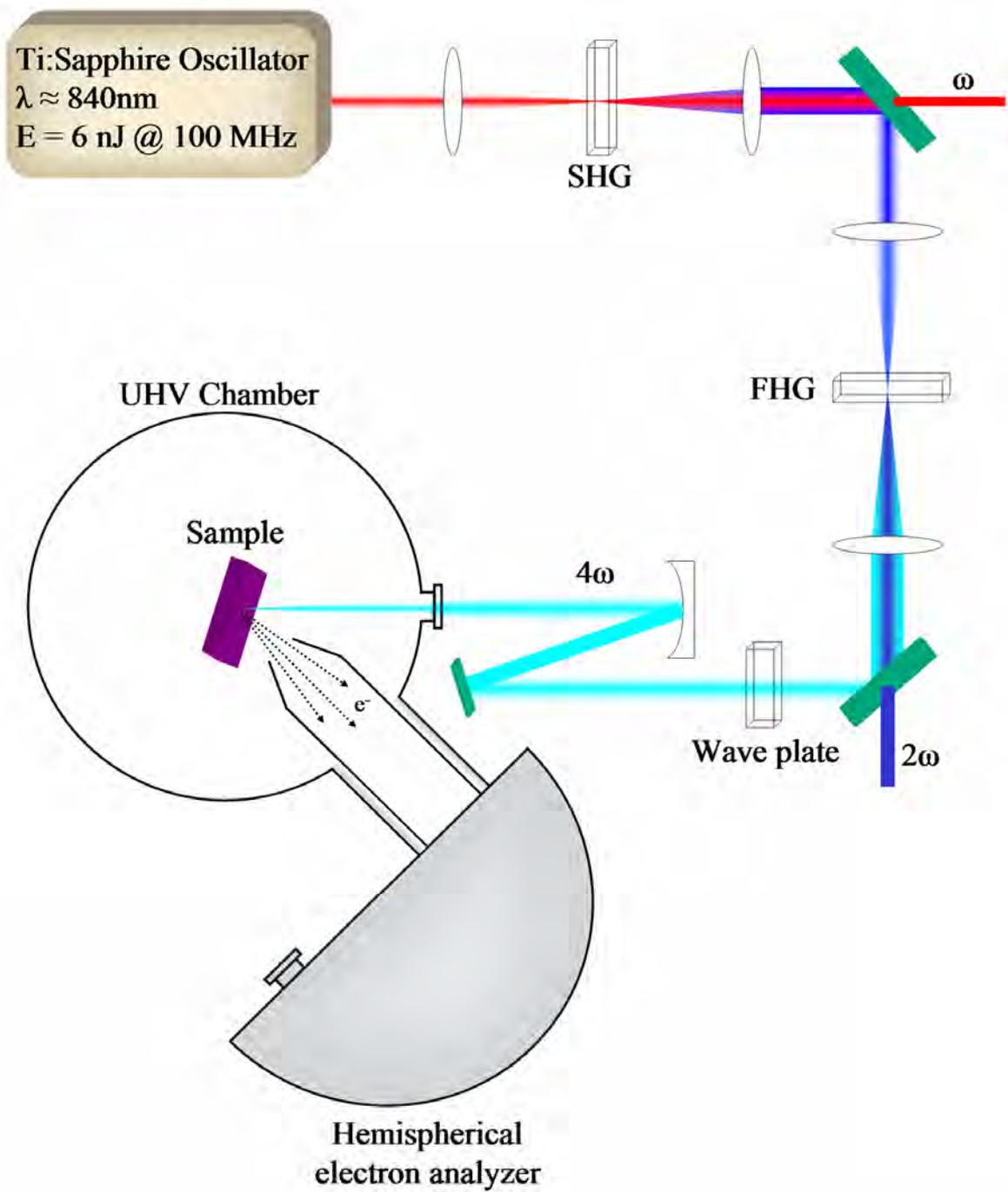

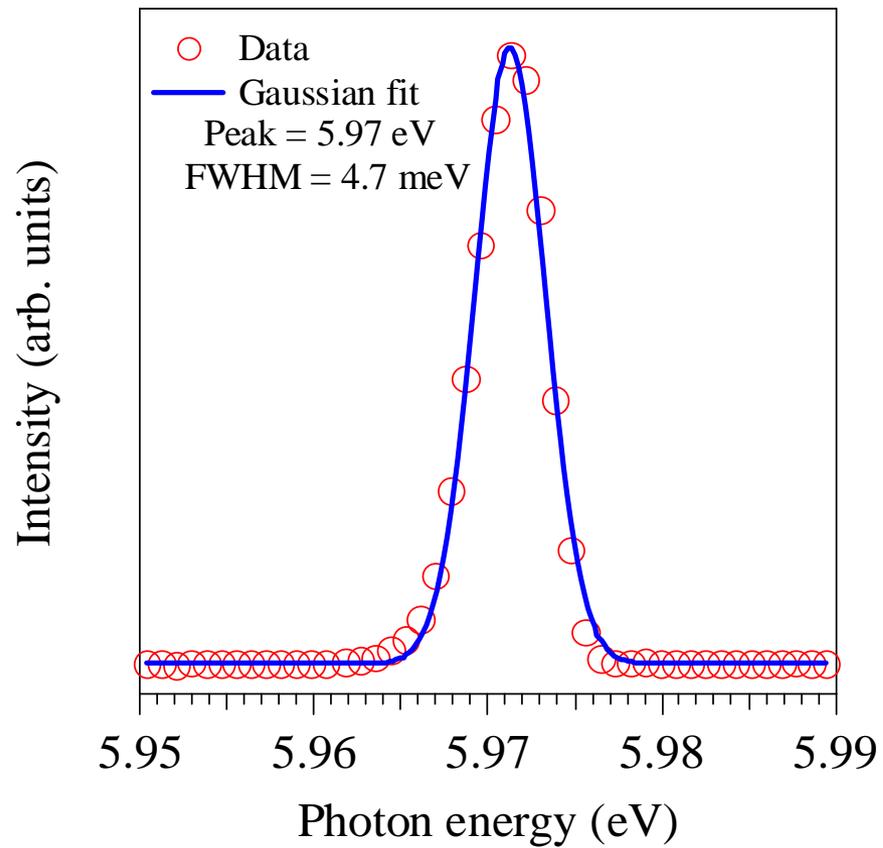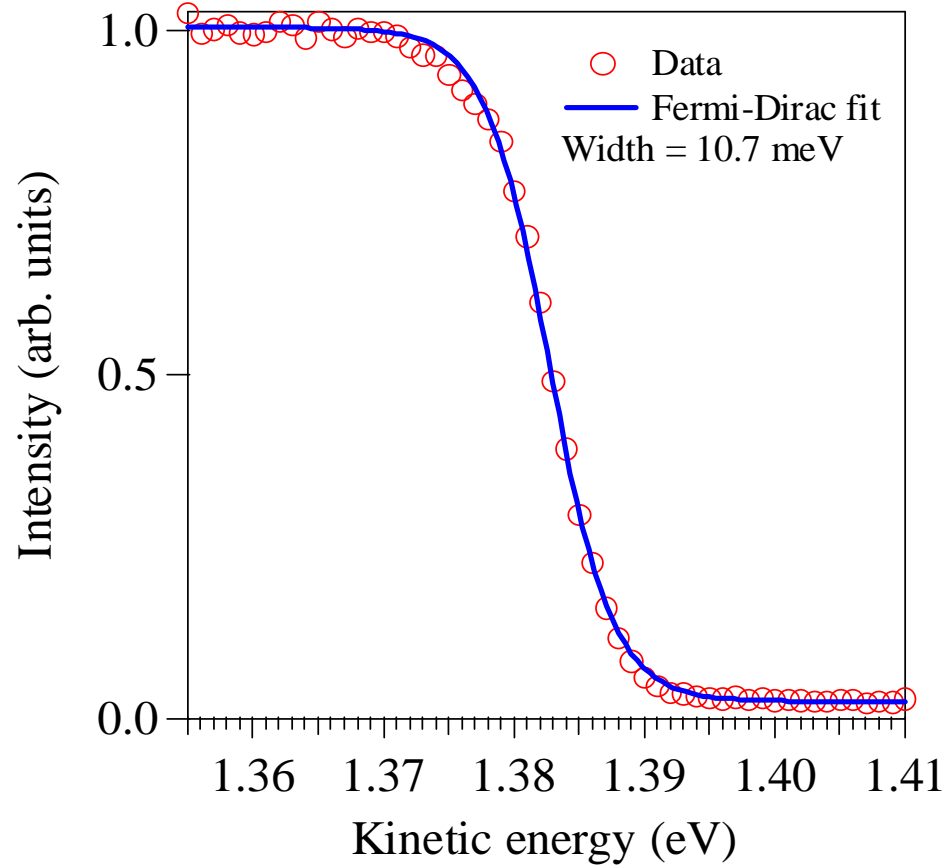

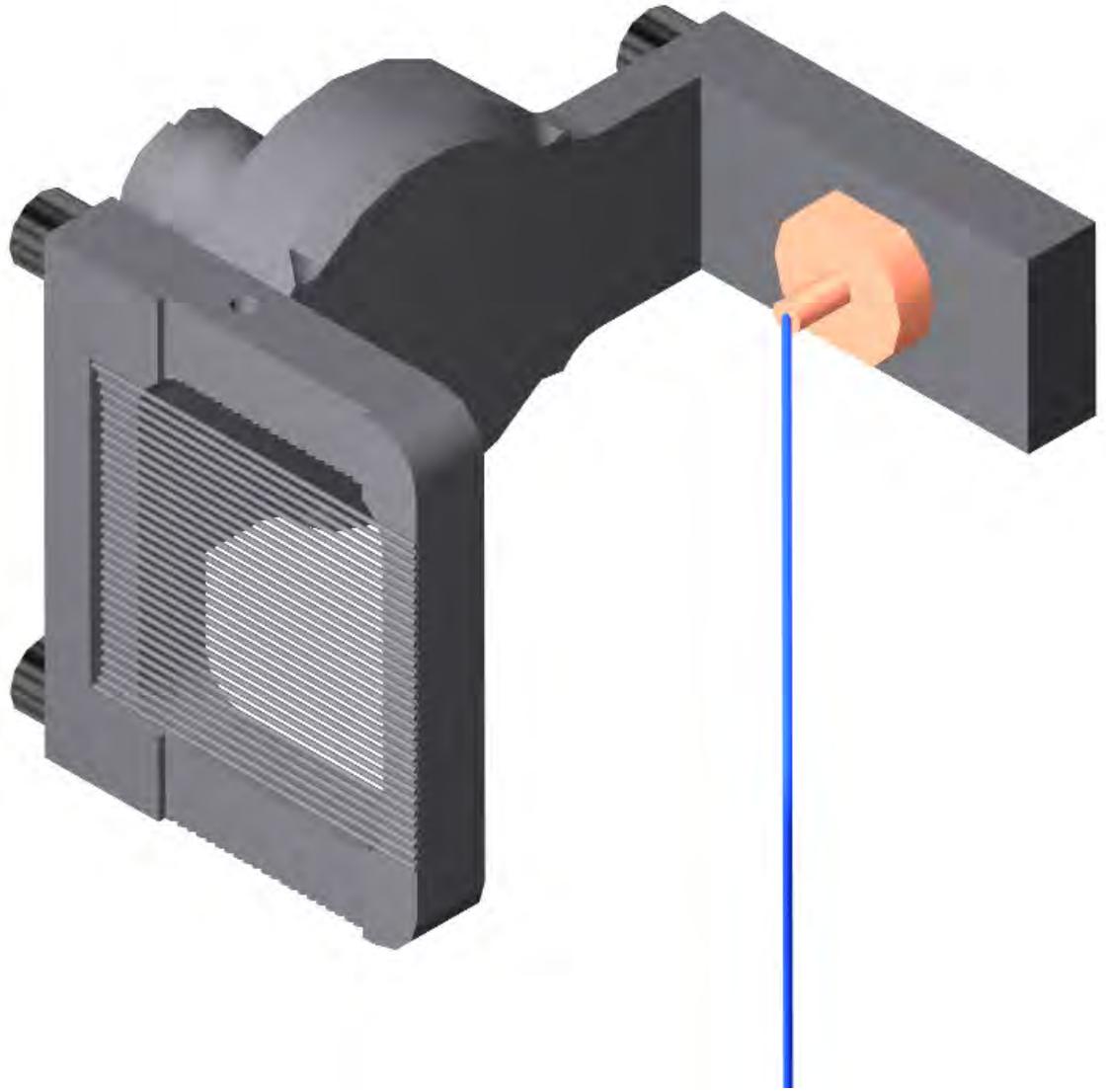

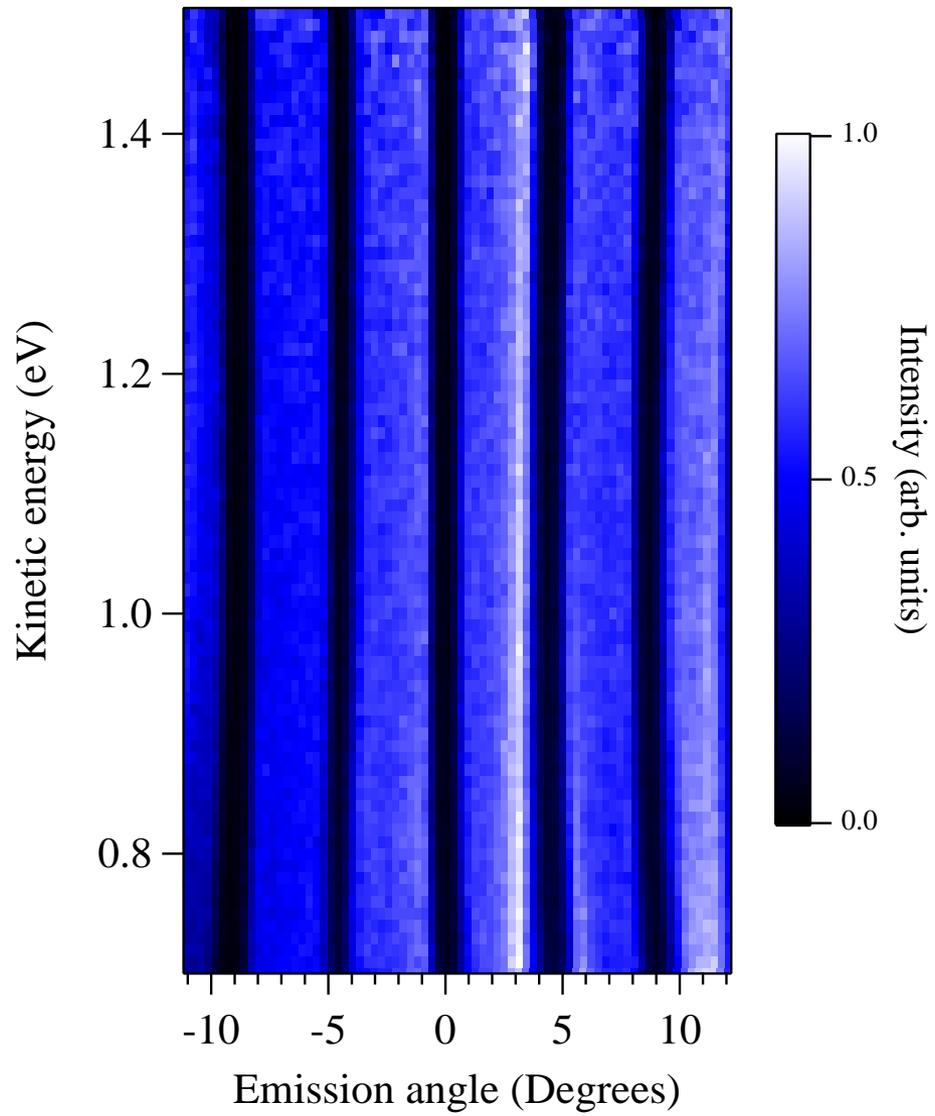

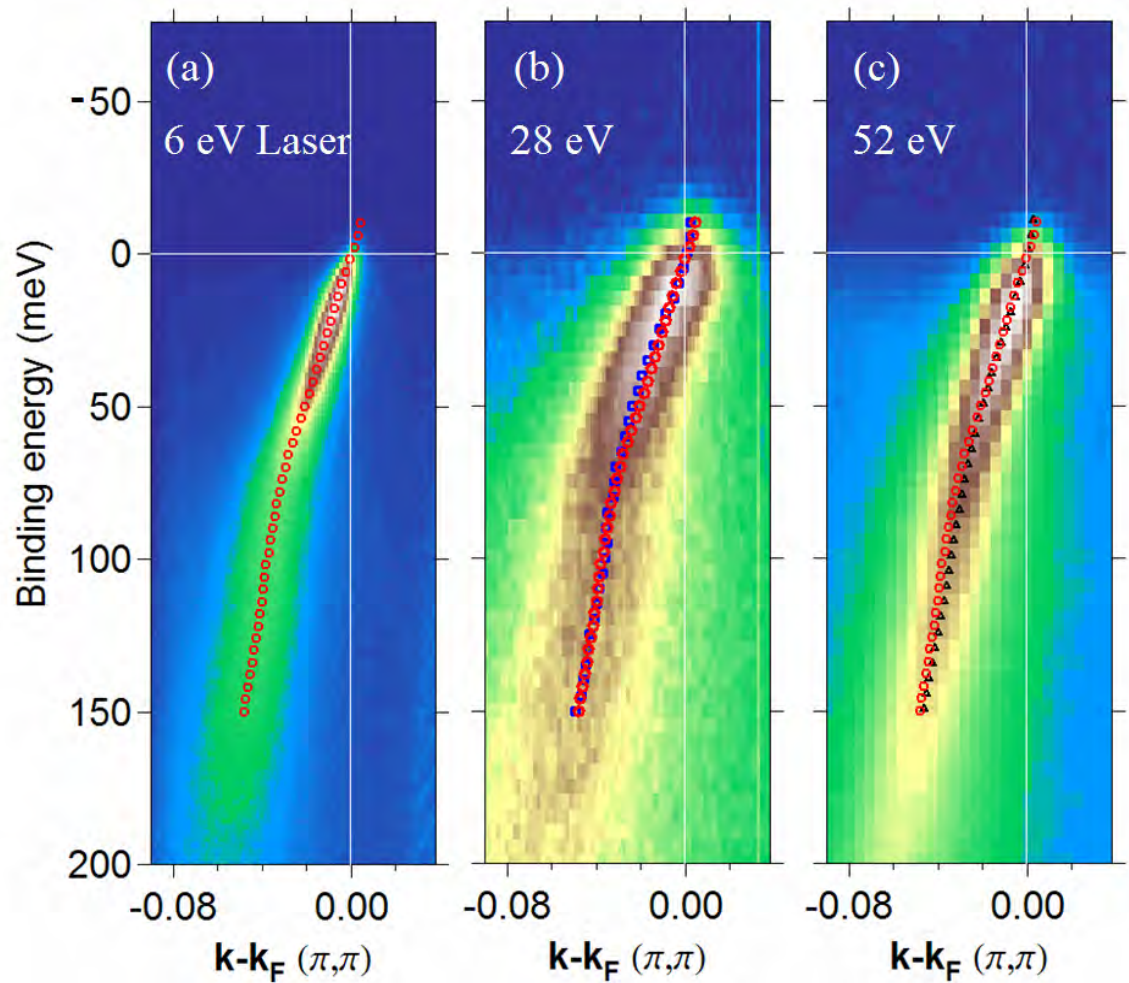

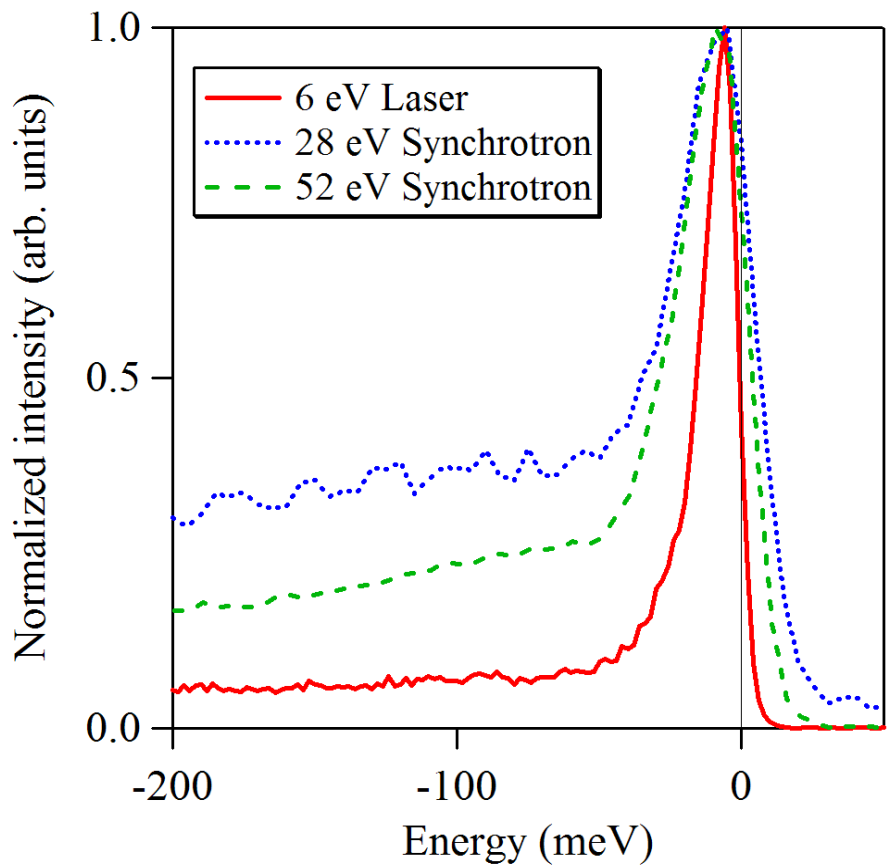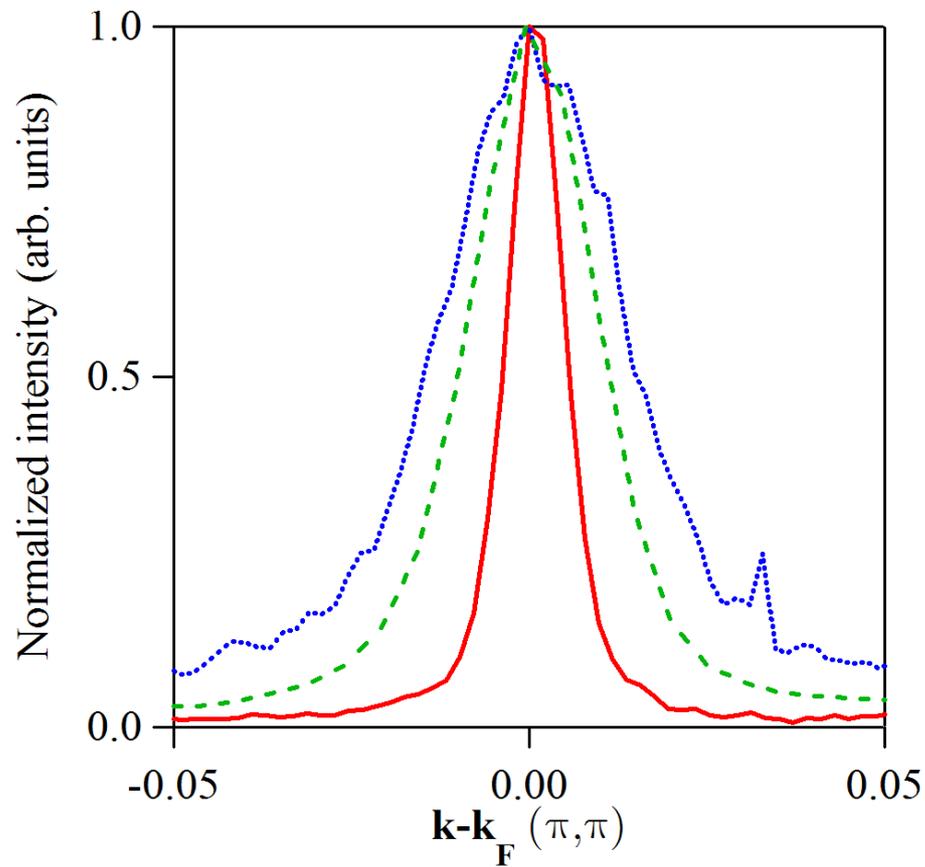